\documentclass[twocolumn,aps,prl,superscriptaddress,amssymb]{revtex4}
\usepackage{setspace}
\usepackage{subfigure}
\usepackage{graphicx,epsf,epsfig}
\usepackage{dcolumn}
\usepackage{bm}
\usepackage{ulem}
\usepackage{color}

\begin{document}
\title{Coupling optical and electrical gating for electronic read-out of quantum dot dynamics}
\author{Smitha Vasudevan}
\affiliation{Department of Electrical and Computer Engineering \\
University of Virginia, Charlottesville, VA 22904 \\}
\author{Kamil Walczak}
\affiliation{Naval Research Laboratory\\
Electronics Science $\&$ Technology Division, Washington, DC 20375\\}
\author{Avik W. Ghosh}
\affiliation{Department of Electrical and Computer Engineering \\
University of Virginia, Charlottesville, VA 22904 \\}
\begin{abstract}
We explore the coherent transfer of electronic signatures from a strongly correlated, optically gated nanoscale quantum dot to a weakly interacting, electrically backgated microscale channel.
In this unique side-coupled `T' geometry for transport, we predict a novel mechanism for detecting Rabi oscillations induced in the dot through quantum, rather than electrostatic means. This detection shows up as a field-tunable split in the Fano lineshape arising due to interference between the dipole coupled dot states and the channel continuum. The split is further modified by the Coulomb interactions within the dot that influence the detuning of the Rabi oscillations. Furthermore, time-resolving the signal we see clear beats when the Rabi frequencies approach the intrinsic Bohr frequencies in the dot. Capturing these coupled dynamics, including memory effects and quantum interference in the channel and the many-body effects in the dot requires coupling a Fock-space master equation for the dot dynamics with the phase-coherent, non-Markovian
time-dependent non-equilibrium Green's function (TDNEGF) transport formalism in the channel through a properly evaluated self-energy and a Coulomb integral. The strength of the interactions can further be modulated using a backgate that controls the degree of hybridization and charge polarization at the transistor surface.

\end{abstract}
\maketitle
Future electronic devices are likely to contain nanoscale components ranging from
random traps, defects and dopants, to engineered memories, logic elements and sensors.
For practical reasons, these elements must interface with a larger microsystem, such
as contacts and substrates. The rapidly increasing
surface sensitivity will make ultimately scaled device properties strongly
dependent on the dynamics and low-frequency noise generated by these nanoscale
components \cite{rucla}. The issue of how the nano and micro domains `talk' to
each other lies at the very heart of the operation of tomorrow's electronics.

The formal challenge in modeling dot-channel hybrid systems arises with their underlying evolution equations themselves. We must directly couple  the
time evolution of the second quantized dot operators in their many-body configuration
(Fock) space \cite{rbhasko} with a quantum kinetic formalism for the channel electrons screened with a mean-field potential \cite{rnegf}. Furthermore, we need to explore the new physics that both the
side-coupled geometry and the different physics in the dot and the channel bring.

\begin{figure}[h]
\centerline{\epsfxsize=3.3in\epsfbox{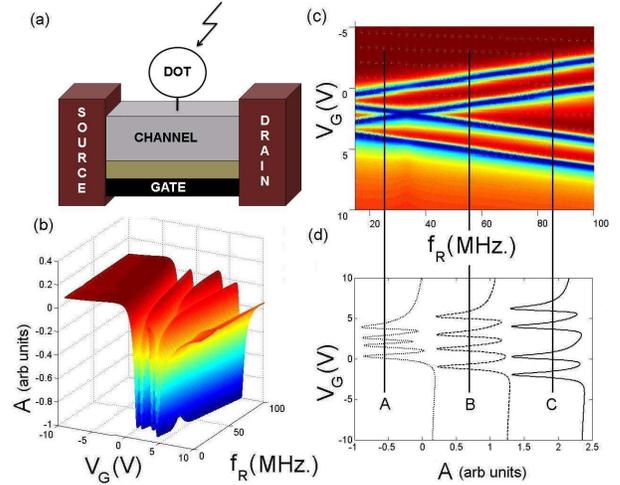}}
\caption{{\it{(a) Electronic detection of Rabi oscillations has relied on Pauli spin blockade between two serially coupled dots \cite{koppens}. In a side-coupled geometry where the dot does not lie on the channel electron's path of propagation, we observe (b) a split of the Fano lineshape proportional to the Rabi frequencies that are tuned by the strength (power) of the incident AC field. (d) The evolving Fano lineshape at 3 different Rabi frequencies. Parameters:
$\epsilon_{0} = 0$ eV, $\epsilon_{1} = 0.1$ eV, $\Delta E_{Z} = 0.2$ meV, $\gamma_{L} = \gamma_{R} = 0.01$ eV, $\tau = 0.25$ eV,
$V_{d} = 0.001$ V.}}}
\label{fanoshape}
\end{figure}

\begin{figure}[h]
\centerline{\epsfxsize=3.3in\epsfbox{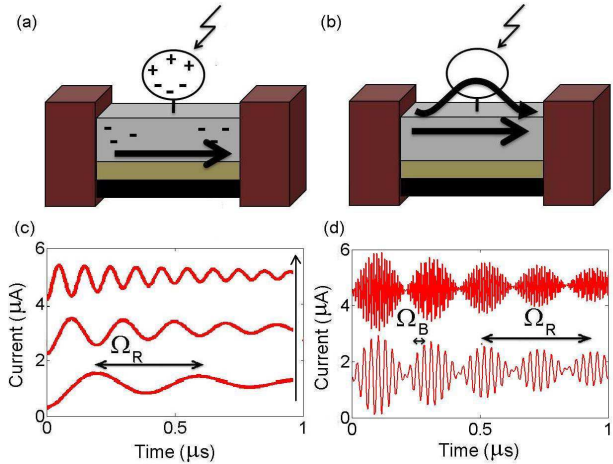}}
\caption{{\it{Schematic description of (a) dipolar Coulomb Scattering (CS) and (b) short-ranged
Quantum Interference (QI) scattering between the dot and channel transport modes.
(c) In the computed channel current for CS, the Rabi oscillations manifest
themselves in the frequency shifts $\Omega_R=2.5 MHz, 5 MHz, 10 MHz$ (in direction of arrow)  ($\Delta\epsilon = 3$ meV). (d) In the QI case, $\Delta\epsilon = 0.2\mu $ eV (e.g. spin split levels in a Zeeman field), $\Omega_R=2.5 MHz$, $\epsilon_{ch} = 0.01$ eV (channel energy), $\epsilon_1 = 0.1$ eV, dot electrochemical potential $\mu = 0$ eV, dot charging energies $ U_1 = 500$ $\mu$eV, $U_2 = 50$ $\mu$eV, contact broadenings $\gamma_L = \gamma_R = 10$ meV, dot-channel coupling $\tau = 0.25$ eV, temperature 100 mK, and decay times $T_1=0.1 \mu$s and $T_2=1 \mu$s. As the Bohr frequency $\Omega_B$ of the levels is comparable with the Rabi frequency, beats are observed in the channel current.}}}
\label{fanocb}
\end{figure}

\begin{figure}[ht]
\centerline{\epsfxsize=3.3in\epsfbox{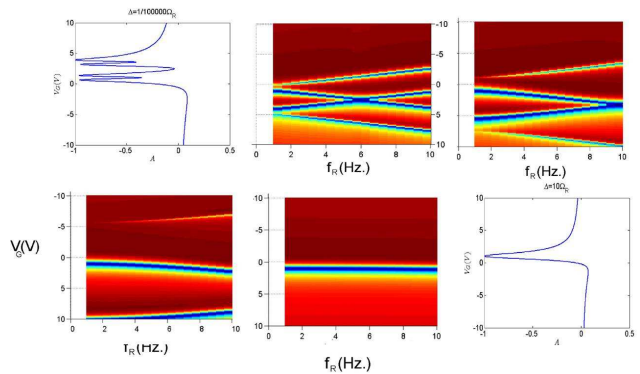}}
\caption{Evolution of Fano lineshape in the presence of Rabi oscillations for varying degrees of intradot Coulomb interactions that detunes the laser frequency with the dot level separation. For unmodified levels we get the Rabi split Fano shapes (Fig.~\ref{fanoshape}), while for strong Coulomb coupling/detuning we register the single level that stays coupled to the channel.}
\label{rfan}
\end{figure}

\begin{figure}[ht]
\centerline{\epsfxsize=2.2in\epsfbox{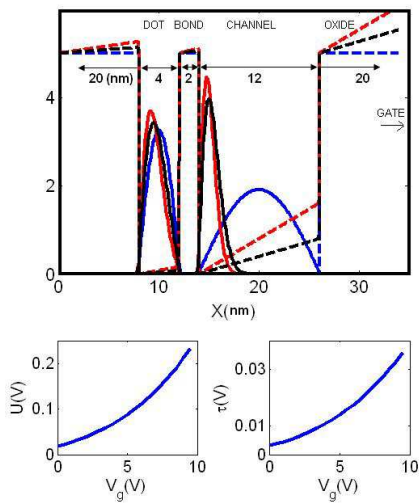}}
\caption{{\it{Top: Band-lineup and eigenstates for dot-channel geometry (dimensions shown) for three representative  gate voltages at $V_{g} = 0V$ (blue line), $2V$ (black line), $3.2V$ (red line) Parameters: $\epsilon_r=1.7$, $\tau_0= 25 \mu eV$ \cite{tauvalue}. The backgate to the right of the figure can be used to tune the electron density and wavefunction at the channel surface, thereby controlling both the Coulomb integral $U$ (bottom left) and the bond coupling parameters $\tau$ (bottom right) between the channel surface and the dot.}}}
\label{qwell}
\end{figure}

In this paper, we (i) present a new theoretical approach to explore the time-dependent interaction between a correlated nanoscale quantum dot and a microscale backgated substrate. The many-electron dynamics of the dot operators are solved to extract one-electron scattering self-energies that drive the simpler TDNEGF channel equations. (ii) The coupling of the dot-channel transport modes creates entirely novel signatures in the channel current spectrum.  A dot coupled to a channel is known to create prominent Fano signatures through gate tunable quantum interference between localized and delocalized states.
 By the incorporation of an added optical gating with a monochromatic laser pulse resonant with the dot level splittings, we can modulate the charge and spin populations and coherences in the dot. This Rabi modulation creates a {\it{split in the Fano spectrum (Fig.~\ref{fanoshape}) that can be tuned with the field strength}} (power of the laser source). (iii) Many-body interactions within the dot further detune this resonant split (Fig.~\ref{rfan}), while (iv) charge depletion/accumulation at the channel surface driven by a back or side gate allows us to control the Coulomb and tunnel couplings between the dot and channel modes (Fig.~\ref{qwell}). Finally, (v) a time-resolved measurement shows prominent {\it{beats}} (Fig.~\ref{fanocb}) in the output signal if the Bohr frequencies of the dot approach the Rabi frequency, as in the case of a dot with two Zeeman split spin states.

{\it{Formalism.}}
Simulating the optical write electronic read process requires a coupling of the Fock-space formalism for
correlated dot dynamics \cite{rbhasko} and the TDNEGF formalism \cite{rjauho2} for weakly interacting quantum channel transport. For a given Hamiltonian, we first solve for the annihilation operators
$c_{1,2}(t)$ using the Heisenberg equation of motion and well-known equal time
anti-commutation rules. While this step can in principle be solved exactly for an isolated dot using exact diagonalization, coupling an interacting electron system with the channel continuum states broadens these
levels through a hybridization procedure that creates a hierarchy of Green's functions, which needs to be truncated at a
suitable point \cite{rmwl}. Assuming weak dot-channel coupling, one can get approximate expressions for these (we will derive explicit forms shortly). Once we have
solved for $c_i(t)$, the dot dynamics can be captured in terms of its retarded and
correlation Green's functions and self-energy matrices
\begin{eqnarray}
g^R_{ij} (t,t')&=& -i\theta (t-t')\langle \{ c_{i}
(t),c^{+}_{j} (t') \} \rangle / \hbar\nonumber\\
g^{n}_{ij}(t,t^\prime) &=& \langle c^{+}_{j} (t') c_{i} (t) \rangle\nonumber\\
\left[\Sigma_s^{(R,in)}(t,t^\prime)\right] &=& \tau\left[g^{(R,n)}(t,t^\prime)\right]\tau^\dagger
\label{eqdot}
\end{eqnarray}
where $\tau = (\tau_1 \tau_2)$ denotes the coupling of the two dot basis states with the channel,
$\langle\ldots\rangle$ denotes a thermal average, $\theta$ is the Heaviside step function, while $\{\ldots,\ldots\}$ denotes
the anticommutator. In addition to these self-energies, responsible for through-bond scattering, quantum interference ({\it{QI mechanism}}) and the exchange of charge between the
dot and the channel,
there is a longer-ranged Coulomb scattering ({\it{CS}}) mechanism that can deplete and polarize channel charges, driven by potential variations $U_1\delta n_1(t) +
U_2\delta n_2(t)$, where the dot charges $n_i(t) = g_{ii}^n(t,t^\prime)$ and the Coulomb
integrals $U_{i} = q^{2}\int dx \int dx' \phi^{*}_{i} (x) \phi_{ch} (x')/ (4\pi\epsilon_{r} |x-x'|)$,
$\epsilon_{r}$ denoting the material dielectric constant and $\phi_{ch}(x)$ denoting the wavefunction in
the channel depth direction x, dictated primarily by the corresponding metal-oxide-semiconductor
electrostatics.The coupling $\tau$ is given by $\tau = \tau_0 \int dx \int dx' \phi^{*}_{i} (x) \phi_{ch} (x')$.
The retarded and in-scattering self-energies and Coulomb potentials are used to compute the transport current in the channel.
In order to do justice to non-Markovian (`memory') effects, we use the full time-dependent non-equilibrium Green's function
(TDNEGF) formalism \cite{rjauho1,rjauho2,rzhu,rhou}.

We start with the retarded and correlation channel Green's functions using the Dyson-Keldysh equations
\begin{eqnarray}
G^R(t,t')&=&g_0(t,t')+\int dt_{1} dt_{2} g_0(t,t_{1}) \Sigma^R (t_{1},t_{2}) G^R(t_{2},t') \nonumber\\
G^{n}(t,t')&=&\int dt_{1} dt_{2} G^R(t,t_{1}) \Sigma^{in} (t_{1},t_{2}) G^{A} (t_{2},t')
\label{eqchannel}
\end{eqnarray}
where $g_0(t,t')$ is the Green's function for the channel decoupled from the dot
(but including Coulomb correlations), $G^A = (G^R)^\dagger$, and $\Sigma^{R,in} =
\Sigma^{R,in}_L + \Sigma^{R,in}_R + \Sigma^{R,in}_s$. For contacts with
broadenings $\Gamma_{L,R}(E)$ and Fermi functions $f_{L,R}(E)$, we can write $\Sigma^{R,in}_{L,R}(t,t^\prime)$ as
the Fourier transforms of $\Sigma^{in}_{L,R}(E) = \Gamma_{L,R}(E)f_{L,R}(E)$ and
$\Sigma^R_{L,R}(E) = -i\Gamma_{L,R}(E)/2 + {\cal{H}}(\Gamma_{L,R}(E))$, where
${\cal{H}}$ denotes the Hilbert transform. From the computed Green's functions and
self-energies, we can now calculate the time-dependent channel current ($\alpha = L,R$)
as
\begin{eqnarray}
I(t) &=& [I_L(t,t)-I_R(t,t)]/2\nonumber\\
I_{\alpha}(t,t^\prime) &=& I^{in}_\alpha(t,t^\prime) - I^{out}_\alpha(t,t^\prime)\nonumber\\
I^{in}_{\alpha} (t,t^\prime)&=&\frac{2q}{i\hbar} \int dt_{1} Tr \Big[ \Sigma^{in}_{\alpha} (t,t_{1})
G^{A} (t_{1},t^\prime) - h.c. \Big] \nonumber\\
I^{out}_{\alpha} (t,t^\prime)&=&\frac{2q}{i\hbar} \int dt_{1} Tr \Big[ G^{n} (t,t_{1})
\Sigma^{A}_{\alpha} (t_{1},t^\prime) - h.c. \Big]
\label{eqtdnegf}
\end{eqnarray}
with $\Sigma^A = (\Sigma^R)^\dagger$ and h.c. the Hermitian conjugate.

{\it{Eqs.~\ref{eqdot}-\ref{eqtdnegf} couple the many-body quantum
dynamics of the dot with the transport properties of the channel}}. The challenge
then is to set up the interacting dot-channel-lead Hamiltonian and compute the dot
operators $c_{i}(t,t^\prime)$.

Laser-irradiating a two-level dot generates a Hamiltonian $H_{int} =
E(t)d_{\mu}\Big(c_{1}^{+} c_{2} + h.c.\Big)$ in the $\{\phi_{1,2}\}$ dot basis.
The transition dipole moment $d_{\mu}=q\int dx \phi_{1}^{*}(x) x\phi_{2}(x)$ and the
laser electric field
$E(t)=E_{0}\cos(\omega_L t)$. Applying the Heisenberg equation for annihilation
$c_{i}$-operators ($i=1,2$), defining the Rabi frequencies $\Omega_R=E_0
d_{\mu}/\hbar$ and Bohr frequencies $\omega_{21}=(\epsilon_2-\epsilon_1)/\hbar$, and assuming the
laser frequency is near resonance with the two-level system with a small detuning
parameter $\Delta = \hbar(\omega_{L} - \omega_{21})$, we can invoke the rotating wave approximation (RWA)

\begin{eqnarray}
\left(
\begin{array}{c}
c_1(t) \\
c_2(t)
\end{array}
\right)
&=&
\overline{U}(t)
\left(
\begin{array}{c}
c_1 \\
c_2
\end{array}
\right)\nonumber\\
\overline{U}(t) &=& \exp\biggl[{-i(H_0-\hbar\Delta/2)t/\hbar}+ {i\vec{\sigma}\cdot{\hat{n}}\Omega t/2}\biggr]
\label{compactcs}
\end{eqnarray}
where $H_0$ is the isolated dot Hamiltonian with eigenvalues $\epsilon_{1,2}$, $\vec{\sigma}$ is the Pauli spin vector, while the unit vector ${\hat{n}} = (\Omega_R,0,\Delta)/\Omega$ with $\Omega=(\Delta^2+\Omega_R^2)^{1/2}$.
The explicit solutions for the $c_{i}$-operators then lead to the Green's functions for the dot using (Eq.~\ref{eqdot})
\begin{eqnarray}
\left[g^R(t,t')\right]&=&-\frac{i}{\hbar}\theta (t-t') \overline{U}(t) \overline{U}^{+}(t')e^{-(t+t^\prime)/T} \nonumber\\
\left[g^{n}(t,t')\right]&=&\overline{U}(t) \rho \overline{U}^{+}(t')e^{-(t+t^\prime)/T},
\label{compactgs}
\end{eqnarray}
where $\rho$ is the density matrix of the isolated dot with diagonal entries given by the equilibrium occupancies of the dot states. The g's in turn yield the scattering term $\Sigma_s(t,t^\prime)$ for the QI mechanism, and the charge polarization $\delta n$ for the CS mechanism. While we accurately capture the dot-channel hybridization, the incoherent coupling of the dot itself to its environment depends on microscopic details of the underlying mechanism, and are included in Eq.~\ref{compactgs} using a phenomenological decay parameter $T$ amounting to a diagonal relaxation time $T_1$ and an off-diagonal decoherence time $T_2$.

One can build interactions into the quantum dot response functions by standard diagrammatic
techniques. {\it{The strength of our
formalism is the coupling of this interacting dot response function with transport
equations in the underlying channel (Eqs.~1-3)}}. For example, we can include an
on-site Hubbard term $U_d$ on the dot by truncating the
hierarchy of dot equations within a `local moment' approximation, yielding a probability weighted result
\begin{eqnarray}
g^{R,n}_{\sigma,int}(t,t^\prime) \approx \Bigl(1-\langle n_{\bar{\sigma}}(t^\prime)\rangle\Bigr)g^{R,n}_\sigma(t,t^\prime) + \nonumber\\
\langle n_{\bar{\sigma}}(t^\prime)\rangle
g^{R,n}_\sigma(t,t^\prime)e^{-iU_d(t-t^\prime)}
\label{pb}
\end{eqnarray}
for a given spin $\sigma$ and its inverse $\bar{\sigma}$. Higher order terms can generate
further interactions, such as high-temperature Kondo correlations \cite{rmwl}.

{\it{Results: {\it{Influence of Rabi on Fano.}}}}
In the past, electronic detection of Rabi oscillations relied on Pauli spin blockade
between two serially coupled dots \cite{koppens}. In a parallel transport geometry,
however, the channel electrons are not Pauli Blocked by the side coupling as the
dot does not lie on the channel electron's path of propagation. The coupling of paths
can occur through a longer-ranged Coulomb interaction that depletes the channel electrons
(Fig.~\ref{fanocb}a) and generates time-resolved transport signatures reminiscent \cite{koppens} of
spin Blockade (Fig.~\ref{fanocb}c). However, a much more interesting mechanism avails itself in
this specific side-coupled T-geometry, where lateral transport paths through the channel continua and the localized dot states interfere quantum mechanically through direct chemical bonding between the dot and the channel surface atoms (Fig.~\ref{fanocb}b), generating a phase-coherent Fano interference.

While Fano signatures are common in quantum dots\cite{fano} , in the dual optical-electronic gated system described here {\it{the Fano lineshape gets convolved with the Rabi signatures impressed upon the dot by the AC field}}. Simplifying Eq.~\ref{compactgs}, we see that the Rabi-Fano interaction manifests itself in the DC spectrum as a field-tunable split in the Fano lineshape (Fig.~\ref{fanoshape}), and as field-tunable beats in the time-resolved signal (Fig.~\ref{fanocb}d) when the Rabi frequency approaches the intrinsic Bohr frequencies in the dot to which the laser is resonantly tuned. The origins however are patently different -- the spectral modifications in Fig.~\ref{fanoshape} arise from $g^R$, while the temporal beats and memory effects in Fig.~\ref{fanocb} arise from $g^n$. Dipolar Rabi frequencies are typically much smaller than electronic state spacings, but the two do approach each other in their spintronic analog where the tunnel coupling is replaced by the Heisenberg exchange between the spin bits and the channel current \cite{stegner}, and the Zeeman splitting of spin states is often comparable to the ESR frequency. Capturing these beats correctly requires careful attention to memory effects that our TDNEGF model incorporates.

{\it{Parameter dependence and tunability.}} We can readily include Coulomb correlations in our model, say by using Eq.~\ref{pb}. Self-interaction correction will make one level repel the other through a mutual Coulomb potential. If the laser frequency is resonantly tuned to the bare level splitting, increasing Coulomb corrections will increase the detuning $\Delta$. As Fig.~\ref{rfan} suggests, the detuning will spread out the spectral weights of the Rabi-split Fano. The spectral weights of the two Rabi split levels ($\pm$ for $i=1,2$ respectively), and thereby the change in channel spectral function can be obtained to yield the Rabi modulated Fano lineshape (invoking the small $\Delta$ RWA for simplicity)
\begin{eqnarray}
g_{ii}(E) &\approx& \frac{1- {\Delta}/{\Omega}}{E-\epsilon_1+i\gamma_1 \pm \hbar(\Delta + \Omega)/2} \nonumber\\
&+&  \frac{1 + {\Delta}/{\Omega}}{E-\epsilon_1+i\gamma \pm \hbar(\Delta - \Omega)/2}\nonumber\\
\delta A &=& g_{ch}\Biggl[1 - \frac{(q+\xi)^2}{\xi^2+1}\Biggr]
\end{eqnarray}
where $g_{ch}$ is the bare channel Green's function, and $q$ and $\xi$ are the phase angles of the dot (obtained from $g_{ii}$) and the dot-channel coupled system respectively\cite{ghoshreview}.
The increasing detuning will reduce any beats in the time resolved data and lead ultimately to the bare Fano lineshapes between the dot and channel states unmodified by the Rabi oscillations.

Electronic read-out of qubits is crucial for integrating quantum computing paradigms
with a solid-state architecture compatible with present day microelectronics. It also
provides the possibility of real-time detection of the dot's molecular `fingerprints'\cite{molsig}.
One can engineer the couplings at a molecular level through synthetic chemistry or
by gate tuning. The use of multiple gates would allow us to independently scan the
location of the channel/dot states and also control the dot-channel coupling. The
latter depends on the density of electrons near the surface, as well as the overlap of
wavefunctions between the dot and channel. By using a backgate, we could deplete
or enhance the electron density, and also control the wavefunction overlap through a
Stark shift of the channel states, thereby controlling both our $U$ and $\tau$
parameters for CS and QI respectively (Fig.~\ref{qwell}). The combination of tunability of
parameters and coherent quantum interference signatures (Fig.~\ref{fanoshape}),electronic and optical gating will allow us to explore a rich spectrum of dot states imprinted on the channel current.

{\it{Acknowledgments.}}
We would like to thank Keith Williams, Robert Weikle, and Lloyd Harriott for useful discussions. This work was supported by DARPA-AFOSR, NSF-NIRT and NSF-CAREER awards. K.Walczak is grateful to NRC/NRL for support.

\end{document}